\documentstyle [12pt,epsfig]{article} 
\makeatletter
\@addtoreset{equation}{section}
\makeatother

\newcommand{\BS}{\bigskip}
\newcommand{\SECTION}[1]{\BS{\large\section{\bf #1}}}

 \newcommand {\alp}  {\alpha}
\newcommand {\aefmu}  {\alpha^{eff}(\mu)}
\newcommand {\aefmz}  {\alpha^{eff}(M_Z)}
\begin{document}
\begin{titlepage}
\begin{center}
\vspace*{2cm}
{\large \bf
Mass and Coupling Constant Limits from Convergence Conditions 
on the Effective Charge in QED }
\vspace*{1.5cm}
\end{center}
\begin{center}
{\bf J.H.Field }
\end{center}
\begin{center}
{ 
D\'{e}partement de Physique Nucl\'{e}aire et Corpusculaire
 Universit\'{e} de Gen\`{e}ve . 24, quai Ernest-Ansermet
 CH-1211 Gen\`{e}ve 4.
}
\end{center}
\vspace*{2cm}
\begin{abstract}
The massless fermion limit of QED is discussed. For on-shell renormalisation
the high energy behaviour fixes no lower limit on the mass of the lightest
fermion if the fine structure constant $\alpha$ is allowed to vary. The choice
of an arbitary (space-like) subtraction point does, however, fix a lower limit 
on the mass of the lightest fermion, for any subtraction scale $\mu$, if the
effective charge $\aefmu$ respects both quantum mechanical superposition
and renormalisation scale invariance. Limits on the values of $\alpha$ or
the electron mass are obtained within the Standard Electroweak Model
by requiring convergence of $\aefmz$.
\end{abstract}
\vspace*{1cm}
\newline
{\it Keywords ;} Quantum Electrodynamics, Fermion Mass Singularities,
Renormalisation Group Invariance,
Standard Electroweak Model.
\newline

\end{titlepage}

\SECTION{\bf{Introduction}}
The massless fermion limit of QED has been discussed in the classic papers of
Kinoshita~\cite{x1} and Lee and Nauenberg~\cite{x2}. The results contained
in these papers are traditionally referred to in the literature as the `KLN'
theorem.
\par The essential conclusion of Kinoshita was the remark that, although 
unrenormalised Ultra-Violet (UV) divergent amplitudes containing fermion 
loops are finite in the massless fermion limit, this is no longer the case
after charge renormalisation. The renormalised amplitudes contain terms 
proportional to $\alp^2\ln(Q/m_f)$ where $\alp$ is the fine structure constant, 
$Q$ the physical scale and $m_f$ the renormalised fermion mass. For fixed $\alp$, $Q$ 
the amplitudes are then logarithmically divergent as $m_f \rightarrow 0$. If, however,
$\alp$ and $m_f$ are treated as free parameters, then, as discussed below, with a 
suitable choice for $\alp$ the divergent part of the amplitude may have any
 finite value, for any non-zero value of $m_f$, however small. Indeed, experimental
 measurements at scales $\gg m_f$ cannot then separately determine $\alp$ and $m_f$.
 Since, in this case, $m_f$ can, in practice, be as small as desired, such a theory
 is called here `quasi-massless'.
 \par Lee and Nauenberg~\cite{x2} discussed the collinear mass singularities related
 to the radiation of real photons from external photon lines. They showed, by 
 considering suitable sums of `degenerate' processes (where a fermion line is
  indistinguishable from a fermion and a collinear photon), that the mass singularities
  cancel. In the corresponding Feynman diagram calculations, mass singular 
  logarithms appear at intermediate stages of the calculation but cancel, order-by-order
  in $\alp$, when real and virtual contributions are added~\cite{x3}. In the case that
  the real and virtual diagrams for final state radiation form a gauge invariant set,
  the cancellation of logarithms is exact. The cancellation does not, however, occur
  for initial state radiation, or if cuts are applied to the angles and energies of
  the final state photons. 
  \par The KLN theorem has been discussed in some detail in a recent paper by the
   present author~\cite{x4} where it is pointed out that uncancelled mass-singular
   logarithms occur at $O(\alp^2)$ in final state radiative corrections due to 
   diagrams where a virtual photon splits into a fermion pair. The corresponding
   renormalised amplitudes are related via analytical continuation and unitarity
   cuts to the UV divergent ones discussed by Kinoshita.
   \par In the present paper radiative corrections due to vacuum polarisation loops
   (forming a gauge invariant set) in the renormalised amplitude for the scattering
   of unequal mass fermions by one-photon exchange is considered. A similar process
   ($e$-$\mu$ scattering) was also considered in Ref.[4], using a number of different
   renormalisation schemes, and where the mass-singular nature of the amplitude for
   any non-vanishing value of $\alp$, in accordance with the conclusions of Ref.[1],
   was confirmed. Here the `quasi-massless' limit of the theory, where $\alp$ and $m_f$
    are treated as free parameters, will be considered, as well as constraints derived
   by requiring convergence of the Dyson sum of the one-loop effective charge for an
   arbitary renormalisation scale. 
   \par The plan of the paper is as follows. In the following Section the information
   on the value of $\alp$ and the fermion masses that may be derived from ideal
   measurements of the differential cross-section of the lepton-lepton scattering
   process will be considered. In Section 3 the one-loop Renormalisation Group
   Equation (RGE) for the effective charge $\alpha^{eff}$ is recalled, as well as
   the relation of its solution
    to the diagrammatic description of the lepton-lepton scattering
   amplitude, taking into account radiative corrections due to vacuum polarisation
   insertions. Requiring convergence of the Dyson sum will be shown to give an upper 
   limit on the renormalisation scale for which the RGE is valid, and also a
   lower limit on the mass of the lightest fermion. Theories of the `quasi-massless'
   type are thus excluded if the solution of the RGE is convergent. In Section 4 the
   phenomenology of the formalism of Section 3 is developed and combined with the
   Standard Electroweak Model to provide bounds on the fine structure constant
   or the mass of the lightest fermion.
    
\SECTION{\bf{Constraints on $\alpha$ and the Fermion Masses
from Measurements of the Effective Charge}}
 The renormalised invariant amplitude for unequal mass charged lepton-lepton
 scattering, including, to all orders, the effect of one-loop vacuum
 polarisation insertions: ${\cal M}^{\infty}(Q)$, is related to the effective
 charge $\alpha^{eff}(Q)$ by the expression:
 \begin{equation}
 {\cal M}^{\infty}(Q) = {\cal M}^{0}(Q) \alpha^{eff}(Q) / \alpha
 \end{equation}
 where ${\cal M}^{0}(Q)$ is the Born-level amplitude, $\alpha = \alpha^{eff}(0)
 = 1/137.036...$, and $Q$ is the physical scale of the scattering process. 
  Since, at the Born level, only a single diagram contributes to the scattering
  process, Q is unambiguously defined by the (space-like) virtuality of the
  exchanged photon (or photons), which is the Mandelstam variable $t$:
  \begin{equation}
  Q^2 = -t = (P_A-P_B)^2\simeq \frac{s}{2}(1-\cos \theta^{\star})
  \end{equation}
  In Eqn.(2.2), $\sqrt{s}$ is the total 
  centre-of-mass energy, $\theta^{\star}$ is the
  centre-of-mass scattering angle, and the 4-vectors
   $P_A$,$P_B$ are defined in Fig.1, 
  which also shows the diagrammatic representation of Eqn.(2.1). The last 
  member of Eqn.(2.2) is valid for $ s \gg m_1^2, m_2^2$ where $m_1,m_2$
  are the masses of the leptons $l_1$ and $l_2$. The effect of including
  Fermion Vacuum Polarisation Loops (FVPL) to all orders is to
  replace the `bare' photon propagator of the Born term by the dressed
  propagator, indicated in Fig.1 by the solid vertical line.
  The renormalisation scale dependent electric charge of the Born term
  (represented in Fig.1 by open squares) is replaced by the effective
  charge at the physical scale $Q$: $e^{eff}(Q)$ (the solid black
  circles) related to $\alpha^{eff}(Q)$ by:
   \begin{equation}
   \alpha^{eff}(Q) = e^{eff}(Q)^2/ 4 \pi
   \end{equation}
   In Eqn.(2.1) the conventional on-shell renormalisation scheme has been 
   used, so that the coupling constant appearing in the Born term is
   just $\alpha$. The infinite sequence of diagrams shown in Fig.1 gives a
   geometric series for the effective charge. Using on-shell
   renormalisation:
   \begin{eqnarray}
   \alpha^{eff}(Q) & = & \alpha [ 1 + \Pi_{(Q)}^{(1)} + (\Pi_{(Q)}^{(1)})^2+...]
   \nonumber \\
           & = & \frac{\alpha}{1-\Pi_{(Q)}^{(1)}} 
  \end{eqnarray}
    Here $\Pi_{(Q)}^{(1)}$ is the one-loop (indicated by the superscript) Photon
    Proper Self Energy Function (PPSEF) in the on-shell scheme. For $n_f$
   different species of fermions \footnote{ For quarks the effect of the colour 
   quantum number of QCD can be correctly included by assigning quarks
   of different colours to different species.}, of mass $m_i$ and charge (in units of that
   of the positron) $q_i$, the PPSEF is given by the expression:
  \begin{equation}
  \Pi_{(Q)}^{(1)} = \frac{2 \alpha}{3 \pi} \sum_{i=1}^{n_f} q_i^2 {\cal L}(\rho_i)
  \end{equation} where $\rho_i = Q^2/4 m_i^2$ and~\cite{x5}
  \begin{equation}    
  {\cal L}(\rho) \equiv \frac{1}{2}(1-\frac{1}{2 \rho}) \sqrt{1+\frac{1}{\rho}}
  \ln \left\{  2 \sqrt{\rho}[\sqrt{\rho}+\sqrt{1+\rho}]
  +1 \right\}+\frac{1}{2 \rho} 
  -\frac{5}{6}
  \end{equation}
  For $\rho_i \gg 1$ the function ${\cal L}(\rho_i)$ reduces to the simplified 
  asymptotic form:
\begin{equation}
{\cal L}^{\infty}(\rho_i) = \frac{1}{2} \ln 4 \rho_i -\frac{5}{6}
 = \ln \frac{Q}{m_i}-\frac{5}{6} 
\end{equation} 
On the other hand, for $\rho_i \ll 1$,  ${\cal L}(\rho_i) \rightarrow 0$,
so that, at low scales, the contributions of heavy fermions to the effective
charge decouple.
\par The contribution of leptons with $|q_i|=1$ to the PPSEF is first 
considered. If there is a hierarchy in the lepton masses such that:
\[ m_1 \ll m_2 \ll m_3~....  \]
then, away from the `threshold'
\footnote{Since the 4-momenta squared of the virtual photons
are space-like there is no actual threshold, but, because of the 
decoupling property of the function ${\cal L}$, the functional 
dependence of $\alpha^{eff}$ on $Q$ changes when $Q \simeq m_i$.}
 regions, where $Q \simeq m_i$, the effective charge is described by the
 following simple formula:
 \begin{equation}
 \frac{1}{\alpha^{eff}(Q)}= -\frac{2 n_f}{3 \pi} \ln \frac{Q}{Q_L(n_f)}
 \end{equation}
 where
 \begin{equation}
 Q_L(n_f) \equiv \overline{m}_{n_f} \exp \left[\frac{3 \pi}{2 n_f \alpha}+
 \frac{5}{6}\right]
 \end{equation}
 and 
 \[  \overline{m}_{n_f} = \left(m_1 m_2 ... m_{n_f} \right)^{\frac{1}{n_f}}  \]
 Here, $n_f$ is the number of active lepton flavours. If $m_i \gg Q$, the
 lepton $i$ is decoupled and therefore `inactive' at the scale $Q$. The evolution of
  $[\alpha^{eff}(Q)]^{-1}$ for $0 \le Q \le m_{\tau}$ , according to Eqn.(2.8)
  is shown in Fig.2. For $Q \simeq m_e, m_{\mu}, m_{\tau}$ the approximation 
  used in Eqn.(2.8) breaks down and the full expression Eqns.(2.4)-(2.6) for
  the effective charge should be used. The subscript `$L$' on $Q$ in Eqn.(2.9)
  stands for `Landau', and it can be seen from Eqn.(2.8) that, when $Q = Q_L$,
  the effective charge becomes infinite~\cite{x6}. As shown in Fig.2,
  the larger the number of active lepton flavours, the smaller is $Q_L$.
  The Landau scales for 1, 2, 3 active flavours are given by the intersections 
  of the lines L1,L2,L3 with the abscissa $[\alpha^{eff}]^{-1} = 0$.
  The respective scales are: $10^{277}$, $10^{138}$, $10^{93}$ GeV, which may be
  compared to the Planck scale $Q_P$ of 1.22$\times$10$^{19}$ GeV.
  These `supra-cosomological' scales are due to the appearance of 
  $(\alpha)^{-1}$ in the exponential factor in Eqn.(2.9).
  \par The question is now asked:`What information on the
   values of the lepton masses and $\alpha$
   can be derived from measurements of $\alpha^{eff}(Q)$?'
   A semi-realistic gedankenexperiment could simply measure the differential
   cross-section $d \sigma/ d t$ for the lepton-lepton scattering process
   shown in Fig.1. Applying all QED radiative corrections, except those due to
   FVPL, to $d \sigma/ d t$ yields $d \sigma^{corr}/ d t$. The effective
   charge is then given by the expression:
   \begin{equation}
   \alpha^{eff}(t) = \left[ \frac{1}{2 \pi} \frac{s^2 t^2}{s^2+u^2}
   \frac{d \sigma^{corr}}{dt}\right]^{\frac{1}{2}}
   \end{equation}
 where $u$ is the third Mandelstam variable.
   Noting that $Q^2 = -t$, then the masses of the $\mu$ and $\tau$, in the
   example shown in Fig.2, can be determined by observing the positions of the
  `kinks' in the logarithmic evolution of $[\alpha^{eff}(Q)]^{-1}$. In fact, it is
  sufficient to derive the Landau scales $Q_L(1)$, $Q_L(2)$, $Q_L(3)$ from the 
  lines L1,L2,L3 in Fig.2 to determine $m_{\mu}$ and $m_{\tau}$.
  Eqns.(2.9),(2.10) give: 
\begin{eqnarray}
m_{\mu} & = & \frac{Q_L(2)^2}{Q_L(1)} \exp \left(-\frac{5}{6}\right) \\
m_{\tau} & = & \frac{Q_L(3)^3}{Q_L(2)^2} \exp \left(-\frac{5}{6}\right) \\
\end{eqnarray}
Unless the experimental resolution is sufficiently good to directly observe 
the decoupling of the electron\footnote{ For scales below the electron mass,
the effective charge is essentially constant, and the classical (Thomson)
limit is approached. This region is discusssed in detail in Ref.[7]}
the values of $m_e$ (the mass of the lightest fermion) and $\alpha$ cannot
be separately determined. Their values are, however, constrained by the
equation:
\begin{equation}
Q_L(1) = m_e \exp \left(\frac{3 \pi}{2 \alpha}+\frac{5}{6}\right) 
\end{equation}
   Some value of $\alpha$ can reproduce the measured value of $Q_L(1)$ for 
   any value of $m_e$, however small. Such a `quasi-massless' theory is 
   indistinguishable from conventional QED for measurements of the effective
   charge, based on Eqn.(2.10), and sensitive only to scales $Q \gg m_e$.
   Of course, many low energy experimental measurements can easily distinguish
   between conventional QED and a quasi-massless version with much smaller
   values of $m_e$ and $\alpha$. For example, the energy levels of the hydrogen
   atom are $\simeq m_e \alpha^2$. However, the the impact of the actual 
   values of the particle masses on the high energy behaviour of a theory 
   remains of great interest, particularly in view of a similar discussion
   for QCD, where, because of confinement, no classical limit can be defined.
   
\SECTION{\bf{Convergence Properties of the Effective Charge for 
Arbitary Renormalisation Scales. Renormalisation Group Equations}}
  Making the replacement $Q \rightarrow \mu$ in Eqns.(2.4)-(2.6), and then 
  eliminating the fine structure constant $\alpha$ between the new equations and
  (2.4)-(2.6), leads to an alternative expression for the effective charge
  $\alpha^{eff}(Q)$:
\begin{equation}
\alpha^{eff}(Q) = \frac{\alpha^{eff}(\mu)}
{1-\frac{2\alpha^{eff}(\mu)}{3 \pi} \sum_{i=1}^{n_f}q_i^2
[ {\cal L}(\rho_i)-{\cal L}(\tilde{\rho}_i)]} 
\end{equation}
where 
\[ \tilde{\rho}_i \equiv \frac{\mu^2}{4 m_i^2} \]    
Since this equation holds for all values of $\mu$, (modulo the convergence
constraints to be discussed below), the right side of Eqn.(3.1) is independent
of $\mu$. Setting $\mu = 0$, for example, Eqns.(2.4)-(2.6) are recovered, as 
${\cal L}(\tilde{\rho}_i) = {\cal L}(0) = 0$. It is convenient, for the 
subsequent discussion, to set $Q = 0$
\footnote{Actually, $Q$ does not need to be strictly zero, but to statisfy
the condition: $ Q \ll m_1$ where $m_1$ where is the mass of the lightest
fermion, so that, to high accuracy, ${\cal L}(\rho_i) = 0$.}
and to leave $\mu$ as a free parameter. Since $\alpha^{eff}(0) = \alpha$,
Eqn.(3.1) then gives:
\begin{equation}
\alpha = \frac{\alpha^{eff}(\mu)}
{1+\frac{2\alpha^{eff}(\mu)}{3 \pi} \sum_{i=1}^{n_f}q_i^2
{\cal L}(\tilde{\rho}_i)} 
\end{equation}  
What is the physical interpretation of this equation?
Referring to Fig. 1 and Eqn.(2.4), the right side of Eqn.(3.2) represents
the sum of the series of diagrams with FVPL shown in Fig. 1, but with a
different choice of renormalisation subtraction scale; $\mu$ rather than
zero as in Eqn.(2.4). Thus the open square boxes in Fig. 1 now represent 
$e^{eff}(\mu)$ rather than $e^{eff}(0) = e$ as in Eqn.(2.4). Now, Eqn.(3.2)
will be consistent with the diagrammatic description provided that the
denominator of the right side correctly represents the geometric sum of
the Dyson series~\cite{x8}:
\begin{equation}
\alpha = \frac{\alpha^{eff}(\mu)}
{1+\Pi(\mu)} = \alpha^{eff}(\mu) \left[ 1-\Pi(\mu)+ \Pi(\mu)^2+... \right]
\end{equation}
where
\begin{equation}
\Pi(\mu) \equiv \frac{2\alpha^{eff}(\mu)}{3 \pi} \sum_{i=1}^{n_f}q_i^2
{\cal L}(\tilde{\rho}_i) 
\end{equation}
The necessary condition for this is that~\cite{x9}:
\begin{equation} 
  |\Pi(\mu)| < 1
\end{equation}
The convergence condition (3.5) for the geometric series has the following 
three consequences:
\begin{itemize}
\item[(i)] For any renormalisation scale $\mu$ Eqn.(3.2) represents
 a convergent infinite series provided that:
 \[ \alpha^{eff}(\mu)  <  2 \alpha. \]  
\item[(ii)] For any renormalisation scale $\mu$, consistent with the
 condition given in (i), a definite lower limit on the mass of the
 lightest fermion exists.
\item[(iii)] For any given mass of the lightest fermion, an upper limit
 exists for $\mu$ consistent with Eqn.(3.5) and the condition given in (i).
 \end {itemize}
 For simplicity, the case of a single fermion of mass $m$ and charge 
 $|q| = 1$ in the FVPL will be considered. In the subsequent Section the
 contributions of all known leptons and quarks to the effective charge
 will be taken into account. With only one active fermion flavour, 
 and $\rho = \mu^2/4 m^2 \gg 1$ the convergence condition becomes:
\begin{equation}
  \frac{2 \alpha_{\mu}}{3 \pi}\left[ \ln\frac{\mu}{m}-\frac{5}{6}\right] < 1
\end{equation}
where 
\[ \alpha_{\mu} \equiv \alpha^{eff}(\mu) \]
For fixed $\mu$, the lower limit on $m$ is then
\begin{equation}    
m_{MIN} = \mu \exp \left[-\left(\frac{3 \pi}{2 \alpha_{\mu}}+ \frac{5}{6}
\right) \right]
\end{equation}
and for fixed $m$, $\mu$ has the upper limit: 
\begin{equation}    
\mu_{MAX} = m \exp \left[\left(\frac{3 \pi}{2 \alpha_{MAX}}+ \frac{5}{6}
\right) \right]
\end{equation}
Since, when $m = m_{MIN}$ or $\mu = \mu_{MAX}$, $\alpha_{\mu} = 2 \alpha$,
Eqns.(3.7),(3.8) may also be written:
\begin{eqnarray}
m_{MIN} & = & \mu \exp \left[-\left(\frac{3 \pi}{4 \alpha}+ \frac{5}{6}
\right) \right] \\
\mu_{MAX} & = & m \exp \left[\left(\frac{3 \pi}{4 \alpha}+ \frac{5}{6}
\right) \right]    
\end{eqnarray}
The upper limit on the renormalisation scale in Eqn.(3.10), as a function
of the values of fermion mass and $\alpha$, is much stronger than that on
the physical scale $Q$ found by Landau~\cite{x6}. For one active lepton
flavour, Eqn.(2.9) gives, for the Landau scale:
\begin{equation}
 Q_L(1) =  m \exp \left[\left(\frac{3 \pi}{2 \alpha}+ \frac{5}{6}
\right) \right] 
\end{equation}
so that
\begin{equation}
\mu_{MAX}  =  Q_L(1) \exp \left[ -\frac{3 \pi}{4 \alpha} \right] 
  =  Q_L(1) \times 10^{-140}
\end{equation}
Since $Q_L(1)$ for $m = m_e$ is $10^{277}$ GeV, then the corresponding value
of $\mu_{MAX}$ is $10^{137}$ GeV, still a supra-cosmological scale much
 larger than the Planck scale. Setting $\mu = Q_P$
in Eqn.(3.9) gives $m_{MIN} = 10^{-122}$ GeV, which is the comfortable 
factor of $2 \times 10^{-119}$ below the mass of the electron.
Although the convergence condition rigorously excludes theories of the
quasi-massless type, its impact on physics in the interesting range of
scales $m_e < Q < Q_P$ would appear to be negligible. As will be explored
in more detail below, this is no longer the case if all FVPL contributions
are taken into account and $\alpha$ is also treated as a free parameter.
For example, setting $m_{MIN}= m_e$ and $\mu = Q_P$ in Eqn.(3.9) requires
that $\alpha < 0.046$. So this condition requires that the fine structure
constant is `small', no larger than about 6 times its actual value. As shown
in Section 4 below, much more restrictive conditions are obtained by
including all known vacuum polarisation contributions to the effective 
charge.
\par The above considerations may be generalised by choosing an arbitary
scale $Q$ in Eqn.(3.1), rather than setting $Q = 0$. If the scales $Q$, $\mu$
are large compared to all fermion masses the asymptotic form 
${\cal L}^{\infty}$ for ${\cal L}$, Eqn.(2.7), may be used. Then Eqn.(3.1)
(with $q_i = 1$ for all i) may be written as:
\begin{equation}
\alpha_Q = \frac{\alpha_{\mu}}{1+\frac{2 n_f \alpha_{\mu}}{3 \pi}
 \ln \frac{\mu}{Q}} 
\end{equation} 
A convergence condition analagous to Eqn.(3.5) then leads to
 restrictions on the scales $Q$, $\mu$ and the values of $\alpha_{\mu}$
and $\alpha_{Q}$. For $\mu > Q$ these are:
 \begin{itemize}
 \item[a)] for fixed Q:
\begin{eqnarray}
 \mu < \mu_{MAX} & = & Q \exp \left[ \frac{3 \pi}{4  n_f \alpha_Q} \right] \\ 
 \alpha_{\mu} < \alpha_{MAX} & = & 2 \alpha_Q  
\end{eqnarray} 
 \item[b)] for fixed $\mu$:
\begin{eqnarray}
 Q > Q_{MIN} & = & \mu \exp \left[ -\frac{3 \pi}{2  n_f \alpha_{\mu}} \right] \\ 
\alpha_{Q} > \alpha_{MIN} & = & 0.5 \alpha_{\mu} 
\end{eqnarray}  
\end{itemize}
Eq.(3.13) can also be interpreted as a solution of the 1-loop Renormalisation
Group Equation (RGE) for the effective charge~\cite{x10}:
\begin{equation}
Q\frac{\partial a}{\partial Q} = \beta (a) = -ba^2 
\end{equation} 
where 
\[ a \equiv \frac{\alpha^{eff}(Q)}{\pi},~~~b = -\frac{2 n_f}{3}. \] 
Now, while it is always true that Eqn.(3.13) is a solution of Eqn.(3.18),
whether or not $\mu$ and $Q$ satisfy the restrictions (3.14)-(3.17), the
corresponence between the RGE, its solution (3.13), and the geometric sum
shown in Fig.1 (which is simply an expression of quantum mechanical
superposition) only holds if these restrictions are respected. 
To show this, consider the contribution $\alpha_Q^n$ of the first n 
terms of the Dyson sum:
\begin{equation}
\alpha_Q^n \equiv \alpha_{\mu}\left[1-R+R^2-...+R^n\right] = \alpha_{\mu} S_n(R)
\end{equation} 
where the common ratio of the geometric series is:
\begin{equation}
R \equiv -b \frac{\alpha_{\mu}}{\pi} \ln\left(\frac{\mu}{Q}\right) . 
\end{equation}
The finite sum in (3.19) is:
\begin{equation}
S_n(R) \equiv \frac{1}{1+R}- \frac{(-R)^{n+1}}{1+R}
\end{equation} 
while the derivative of $S_n$ with respect to R is:
\begin{equation}
\frac{dS_n}{dR} = -\frac{1}{(1+R)^2}\left\{1-(-R)^n[1+n(1+R)]\right\}
\end{equation}
Eqns.(3.19),(3.20) give:
\begin{equation} 
Q\frac{\partial \alpha_Q^n}{\partial Q} = 
 \frac{\partial \alpha_Q^n}{\partial \ln Q} =
 \alpha_{\mu} \frac{d S_n}{d R}  \frac{\partial R}{\partial \ln Q} =
 b\frac{\alpha_{\mu}^2}{\pi}\frac{d S_n}{d R}.
\end{equation}   
 The values of $S_n(R)$ and $d S_n(R)/dR$, in the large n limit, are shown in
 Table 1 for the five cases:
 \[ R<-1,~~~R=-1,~~~~-1<R<1,~~~R=1,~~~R>1. \]
 The asymptotic ($n \rightarrow \infty$) behaviour of $S_n(R)$ and $d S_n(R)/dR$
  may be read off from the entries in Table 1. For $R \le -1$, 
  $S_n \rightarrow \infty$ and $dS_n/dR \rightarrow -\infty$. For 
  $-1<R<1$ finite, n independent, values $S_{\infty}$ and $d S_{\infty}/dR$
  are found. For $R=1$, $S_n(R)$ and $d S_n(R)/dR$ exhibit finite
  oscillations, while for $R>1$ they undergo infinite (sign alternating)
  oscillations. When $-1<R<1$ (respecting the restrictions (3.14)-(3.17)),
 $S_n(R)$ and $d S_n(R)/dR$ are simply related:
 \begin{equation}
 \frac{d S_n}{d R} = -(S_n)^2
\end{equation}                   
Thus, using Eqn.(3.24), Eqn.(3.23) can be written, in the $n \rightarrow \infty$
limit, as:
\begin{equation} 
Q\frac{\partial \alpha_Q^{\infty}}{\partial Q} = 
 -b\frac{\alpha_{\mu}^2}{\pi}S_{\infty}^2 =
 -\frac{b}{\pi}\left(\alpha_Q^{\infty}\right)^2 
\end{equation}   
which is equivalent to the RGE (3.18). For any other value of $R$, the partial
differential equation (3.23) has no finite limit as $n \rightarrow \infty$
and no RGE is obtained. It may be remarked that only for $-1<R<1$ does the
right side of Eqn.(3.13) correctly represent the infinite sum $S_{\infty}$.
As shown in Table 1, the formula is correct for $-1<R<1$. For $R= -1$,
the correct result, $+\infty$, is also found. However, for $R<-1$, Eqn.(3.13)
gives a finite negative value instead of the  the correct value $+\infty$.
 For $R=1$  the value of 1/2 is found instead of a value oscillating
 between 0 and 1, and, finally, for $R>1$, Eqn.(3.13) gives a finite number
 between 1/2 and 0 instead of the correct (infinitely oscillating) result.
 Only when $R=-1$ (corresponding, in the analagous case of Eqn.(2.4), to
 the Landau singularity) is the breakdown of convergence evident
 from Eqn.(3.13) itself. 
 \par The essential conclusion of this Section is that the formula (3.1) for
 the effective charge corresponds to a finite Dyson sum only for values of
 the renormalisation subtraction scale $\mu$ less than some upper limit.
 This limit is fixed either by the mass of the fermion in the FVPL (Eqn.(3.8))
 or by the physical scale $Q$ (Eqn.(3.14)). For a fixed renormalisation
 scale there are complementary lower limits on the fermion mass (Eqn.(3.9))
 or $Q$ (Eqn.(3.16)). The upper limit on the renormalisation scale 
 determined by the electron mass and $\alpha$ is, although still of
 supra-cosmological magnitude, a factor of $10^{140}$ smaller than the
 Landau scale that fixes the upper limit on $Q$. The RGE for the effective
 charge is only valid within the convergence domain of the renormalisation
 scale, so Renormalisation Group Invariance~\cite{x11} is similarly limited.
 \par In the following Section, constraints on the mass of the
 lightest fermion and the value of $\alpha$ are derived by applying the 
 convergence condition (3.5) to the effective charge, taking into account
 all known 1-loop FVPL contributions.   
   
\SECTION{\bf{Constraints on $\alpha$ and $m_e$ from
Convergence Conditions on the Effective Charge in the Standard
Electroweak Model}}
The convergence condition (3.5) implies that the relation between
the the renormalisation scale $\mu$ and the minimum mass of the lightest 
fermion (i.e. the electron)  $m_{MIN}$, is given by the following expressions:
\begin{eqnarray}
m_{MIN} & = & \mu \exp \left[-\frac{1}{2}\left(\frac{3 \pi}{2 \alpha}
-F(\mu)\right)\right]  \\
F(\mu) & = & 2 \ln\frac{\mu^2}{m_{\mu} m_{\tau}}+2 \ln\frac{\mu}{\overline{m}_d}
+\frac{16}{3} \ln \frac{\mu}{\overline{m}_u}-\frac{100}{9} 
\end{eqnarray}
where
\[  \overline{m}_d = (m_s m_d m_b)^\frac{1}{3},~~~\overline{m}_u
 = (m_u m_c)^\frac{1}{2}   \]
 and 
 \[ m_f \ll \mu < M_W,m_t \]
 so that the top quark contribution and the (gauge dependent) $W$ contributions
 to the effective charge may be neglected.  Eqn.(4.1) is the generalistion,
 including all known fundamental fermion species, of Eqn.(3.9) where only
 one fermion species is included in the FVPL. The values of the fermion mass
 parameters used below in Eqn.(4.2) are presented in Table 2. For each quark
 flavour, 3 species of fermions are included in Eqn.(3.1) to take into 
 account the colour quantum number of QCD. As is common practice~\cite{x12},
 the light quark masses $m_u = m_d, m_s$ are chosen so as to correctly 
 reproduce the non-perturbative hadronic vacuum polarisation contribution 
 deduced, via a dispersion relation, from the experimental data on
 $e^+e^- \rightarrow$ hadrons. The quark and lepton masses in Table 2, 
 when substituted in Eqn.(2.4) give:
 \[ \alpha^{eff}(M_Z)^{-1} = 128.9   \]
 in agreement with recent estimates~\cite{x13}. The Z mass is taken
 to be 91.2 GeV~\cite{x14}.
 \par In the Standard Electroweak Model (SM)~\cite{x15}, $\alpha$ and $M_Z$
 are related by the expression:
 \begin{equation} 
 M_Z^2 = \frac{4 \pi \alpha \upsilon^2}{\sin^2 2 \theta_W}
 \end{equation}
 where $\theta_W$ is the weak mixing angle and $\upsilon$ is the vacuum
 expectation value of the Higgs field. The on-shell renormalisation scheme
 is used, so that:
 \begin{equation} 
 \sin^2 \theta_W = 1-\frac{M_W^2}{M_Z^2}
 \end{equation}
 and virtual electroweak corrections are neglected.
 \par An upper limit, $\alpha_{MAX}$, on the value of the fine structure
 constant is now obtained by requiring that $\alpha^{eff}(M_Z)$ is convergent
 in the sense discussed above
 \footnote{For this it is assumed that the fermion masses and $\alpha$
 enter as uncorrelated parameters in the theory. While this seems reasonable
 for the leptons, which are described purely perturbatively, it is less
 evident for the non-pertubative domain of QCD, described here by
 effective light quark masses. It is possible that changing only  $\alpha$
 while leaving all QCD parameters constant, would modify also the
 effective light quark masses needed to describe the non-perturbative
 domain of QCD. Such refinements are neglected here}.
 The limit obtained depends on the value of the remaining SM parameters:
  $\upsilon$ and $\theta_W$. Two different hypotheses will be made.  
  Firstly, that $\upsilon$ is fixed at its measured value of 
  $\simeq 252$GeV and only $\theta_W$ varies. Since
 \begin{equation}
 G=\frac{1}{\sqrt{2} \upsilon^2}
 \end{equation} 
 where $G$ is the Fermi constant, the strength of the low-energy weak
 interaction remains unchanged in this case. Secondly, the parameter 
 $\upsilon'$ defined as:
 \begin{equation} 
\upsilon'  = \frac{2 \sqrt{\pi} \upsilon}{\sin 2 \theta_W}
 \end{equation}
 is allowed to vary freely.
 Using Eqns.(4.1),(4.3) and (4.6) $\alpha^{eff}(M_Z)$ is found to be 
 convergent provided that $\alpha < \alpha^{MAX}$ where, with 
 $m_{MIN} = m_e$: 
 \begin{equation}  
\frac{\upsilon' x}{m_e}  = \exp \left[\frac{1}{2}\left(\frac{3 \pi}{2 x^2}
-F(\upsilon' x)\right) \right]
 \end{equation}
 and 
 \begin{equation}
 x \equiv (\alpha^{MAX})^\frac{1}{2}
 \end{equation}
 For fixed $\upsilon'$ Eqn.(4.7) is an implict equation for $\alpha^{MAX}$
 that is readily solved numerically. The results are shown, as a function
 of $\sin \theta_W$, for $\upsilon = 252.2$ GeV, in Fig. 3 and also, as a
 function of $\upsilon'$, in Fig. 4. Also shown, in each case, is the value of 
 $M_Z$ given by Eqn.(4.3) when $\alpha = \alpha^{MAX}$. For the experimentally
 measured values of $\sin \theta_W$ and $\upsilon'$ (indicated by the 
 vertical arrow) $\alpha^{MAX}$ is $\simeq 0.05$ (i.e. $\simeq 7 \alpha$)
 and $M_Z^{MAX}$ is $\simeq 250$ GeV. For smaller
 values of $\sin \theta_W$ (or larger values of
 $\upsilon'$) a marginally stronger restriction on $\alpha$ is obtained.
 Note that the minimum value of $\upsilon'$ for constant $\upsilon$
 (and hence the strongest restriction on $M_Z$) is obtained when
$\sin 2 \theta_W = 1$ or $\theta_W = \pi/4$. This limit is indicated by
the hatched vertical line in Fig. 3.
\par Fixing the value of the fine structure constant to its experimental value
 $\alpha^{exp}$, Eqn.(4.7), with the replacement
 $x \rightarrow \sqrt{\alpha^{exp}}$,
 may be solved for the minimum value of the electron
 mass such that $\alpha^{eff}(M_Z)$ is covergent. The results, for different
 values of $M_Z$, are presented in Table 3. The corresponding values of
 $\upsilon' = M_Z/\sqrt{\alpha^{exp}}$ are also shown. At the actual value of 
 the Z mass, corresponding to $\upsilon' = 1068$ GeV,
 the minimum value of the electron mass is 1.9$\times 10^{-128}$ GeV.
It is interesting to note that, when all known fermions are included
in the FVPL, the minimum electron mass, at a scale of 853 GeV, is 
1.1$\times 10^{-120}$ GeV, a factor 100 times greater than the minimum mass
at the Planck scale of 1.22$\times 10^{19}$ GeV, given by Eqn.(3.9), where only
 one fermion species is considered. The convergence conditions become
 very much more restrictive when additional fermion species are included
 in the FVPL.

\SECTION{\bf{Higher Order Corrections}}
 In the on-shell scheme the asymptotic PPSEF for a single fermion species is given,
 up to two loops, by the expression~\cite{x16}: 
 \begin{equation}
 \Pi^{(2)}(Q) = \frac{2 \alpha}{3 \pi}\left[ \ln \left(\frac{Q}{m}\right)
 -\frac{5}{6}\right]
 +\left(\frac{\alpha}{\pi}\right)^2\left[\frac{1}{2}\ln\left(\frac{Q}{m}\right)
 + \zeta(3)-\frac{5}{24}\right]
 \end{equation}
 More generally, the next-to-leading logarithmic contribution to the PPSEF
 at order $\alpha^n$ 
 is~\cite{x17}:
 \begin{equation}
\alpha^n\tilde{\Pi}^{(2n)} =
 \frac{3}{4}\left(\frac{\alpha}{\pi}\right)^n \frac{1}{(n-1)}\left[
 \frac{2}{3} \ln \left(\frac{Q}{m}\right)\right]^{n-1}
 \end{equation}
 The terms given by (5.2) for successive values of $n$ are those of a logarithmic series.
 Thus, the next-to-leading logarithms may be resummed to all orders in $\alpha$ to 
 yield the effective charge, to next-to-leading logarithmic order ~\cite{x18}:
 \begin{equation}
\alpha_{NLL}^{eff}(Q) = \frac{\alpha}{1-\Pi^{(1,\infty)}_{(Q)}+\frac{3 \alpha}{4 \pi}
\ln (1 - \Pi^{(1,\infty)}_{(Q)})}   
 \end{equation}
 where $\Pi^{(1,\infty)}_{(Q)}$ is the asymptotic form of the one loop PPSEF of 
 Eqn.(2.5). 
\par The effect of the next-to-leading logarithmic terms is very small. For example,
 requiring positivity of the effective charge in Eqn.(5.3) leads to the condition:
\[ \Pi^{(1,\infty)}_{(Q)} < 0.9917 \]
instead of the corresponding one-loop condition  $\Pi^{(1,\infty)}_{(Q)} < 1$.
  The Landau scale for one active flavour ($m_1 = m_e$) is reduced from $10^{277}$ GeV to
$10^{275}$ GeV. Similar corrections of $\leq 1 \%$ are expected in the 
value of $\Pi(\mu)$ given by the convergence condition (3.5) due to 
 next-to-leading logarithmic terms. This is much less than other 
sources of error (for, example due to the approximate
 parameterisation of non-perturbative QCD effects by effective quark masses) 
 in the phenomenological discussion of Section 4 above.
In fact, for energy scales of current experimental interest the 
next-to-leading logarithmic terms in Eqn.(5.3) give a simple 
 multiplicative correction to the one-loop PPSEF. Considering, for
 example, one fermion flavour ($m_1 = m_e$) and $Q = 2$ TeV gives 
$\Pi^{(1,\infty)} =0.022$. Then, since for small $x$, $\ln(1-x) \simeq -x$,
the next-to-leading logarithmic correction modifies the one-loop PPSEF
by the constant factor: $1+3\alpha/(4 \pi) = 1.00742$.

\SECTION{\bf{Concluding Remarks}}
This study has shown that, in QED, complementary limits exist bounding from
below the mass of the lightest fermion, and from above the the high energy 
scale at which the theory may be applied, if the freedom of choice of 
subtraction scale, that leads to Renormalisation Group Invariance, is 
respected. For smaller masses and higher scales, the geometric series
(a consequence of quantum mechanical superposition) that yields the 
effective charge is not convergent, and the usual relation between the
Dyson sum and the solution of the RGE breaks down. The limitations found
in this way are much more restrictive than those considered by
Landau~\cite{x6} using the conventional on-shell scheme with a fixed
subtraction point. 

\par The work described here may be extended in two different directions:
\begin{itemize}
\item[(i)] Within QED, consider the effect of multiple fermion loop effects
in the PPSEF, and contributions due to W loops.
\item[(ii)] Apply a similar analysis to the weak and strong interactions.
\end{itemize}
 As discussed in the previous section, even the leading correction
 of the type (i), due to next-to-leading logarithmic corrections,
 is small. 
 On the other hand (ii) may be expected to
 lead to more interesting results. Replacing the invariant amplitude
 for charged lepton scattering considered here by the respective processes:
 $\nu_l l' \rightarrow \nu_l l'$,  $\nu_l l' \rightarrow \nu_{l'} l$ and
 $q q' \rightarrow q q'$, the weak neutral current (Z exchange), weak charged
 current (W exchange) and the strong interaction (gluon exchange) may be
 investigated. In each of these processes, a single diagram contibutes at the Born level
 ,and, as in the QED case considered above, there is a unique physical scale
 in the corresponding `running coupling constant'. The analysis of the
 latter, the analogue, for the weak or strong interactions, of the
 effective charge of QED, is complicated by problems of gauge dependence.
 These have been discussed, for neutral currents in the SM, by Baulieu and
 Coqueraux~\cite{x19}. Because of $\gamma-Z$ mixing effects, it was
 concluded that only for a specific choice of covariant gauge 
 $a = \xi^{-1} = -3$ can effective charges respecting a RGE be associated 
 with the photon and Z propagators.
 \par In the case of QCD, although gluon and quark loops are expected to yield
 a geometric Dyson sum, the gluon loop contribution is gauge dependent, 
 and UV divergent vertex diagrams containing the non-abelian triple gluon 
 coupling must also be taken into account. Qualitatively, the result for the
 running coupling constant is expected to be determined by a geometric sum
 of the type:        
 \begin{equation}
  1-R+R^2-... = \frac{1}{1+R}
 \end{equation}     
 (where R is $> 0$) in contrast to QED where the corresponding
  series has the form:
 \begin{equation}
  1+R+R^2+... = \frac{1}{1-R}
 \end{equation}     
  As is well known, the difference in the sign of $R$ in Eqns.(6.1),(6.2) 
  comes from the dominant contribution, in QCD, of diagrams containing the
  triple gluon coupling, which have the opposite sign (as in the case of
  W-pair loops in QED) to the fermion loops~\cite{x20}.
  Comparing Eqns.(6.1) and (6.2) it can be seen that there is no Landau
  singularity in QCD, where the coupling constant becomes infinite, but that
  a physical limitation will come rather from the convergence of the
  series in Eqn.(6.1), which is of the same type as that discussed above
  in connection with the domain of validity of the Renormalisation
  Group in QED. This problem has been investigated in a companion
  paper~\cite{x21}. In view of the relatively large size of the 
  QCD coupling constant it is to be expected, and indeed found,
  that much more stringent restrictions apply than the Landau
  condition in QED.
 \par {\bf Acknowledgements}
 \par
  W.Beenakker and  M.Consoli are thanked for their interest
 in the work described in this paper and for helpful discussions
 and correspondence.     
\pagebreak
 
\pagebreak
\begin{table}
\begin{center}
\begin{tabular}{|c|c|c|c|c|c|}  \hline
      & $R < -1$ & $R = -1$ & $-1 < R < 1$ & $R=1$ & $R > 1$ \\
\hline
\hline
  & & & & & \\
  $S_n$ & $\frac{-(-R)^{n+1}}{1+R}$ & n+1 & $\frac{1}{1+R}$ & 1 (n even) & 
  $\frac{-(-R)^{n+1}}{1+R}$ \\
  & & & & 0 (n odd) & \\
\hline 
  & & & & & \\
  $\frac{dS_n}{dR}$ & $\frac{n(-R)^{n}}{1+R}$ & $-\frac{n(n+1)}{2}$ 
  & $\frac{-1}{(1+R)^2}$ &  $n/2$ (n even) & 
  $\frac{n(-R)^{n}}{1+R}$ \\
  & & & & $-(n+1)/2$ (n odd) & \\
\hline           
\end{tabular}
\caption[]{ Values of the sum $S_n$ of a geometric series and of $dS_n/dR$, in the 
large $n$ limit, for different values of the common ratio $R$. }   
\end{center}
\end{table}
\begin{table}
\begin{center}
\begin{tabular}{|c|c|c|c|c|c|c|c|c|c|} \hline
Fermion & e & $\mu$ & $\tau$ & u & d & s & c & b & t \\
\hline
\hline
Mass (MeV) & 0.511 & 105.7 & 1777 & 36.9 & 36.9 & 500 & 1570 & 4906
 & 1.8$\times 10^5$ \\
\hline
\end{tabular}
\caption[]{ Fermion mass parameters used for the vacuum polarisation
loops. $\overline{m}_d = (m_s m_d m_b)^\frac{1}{3}$. For $ \mu < m_t$, 
 $\overline{m}_u = (m_u m_c)^\frac{1}{2}$. For $ \ge m_t$,
$\overline{m}_u = (m_u m_c m_t)^\frac{1}{3}$}       
\end{center}
\end{table}
\begin{table}
\begin{center}
\begin{tabular}{|c|c|c|c|c|c|c|} \hline
$M_Z$ (GeV) & 17.1 & 34.2 & 91.3 & 171.0 & 427.6 & 853.2 \\
\hline
\hline
$\upsilon'$ (GeV) & 200 & 400 & 1068 & 2000 & 5000 & 10000 \\
\hline
$m_e^{MIN}$ (GeV) & 5.4$\times10^{-127}$ &  5.5$\times10^{-135}$  & 
1.9$\times10^{-128}$ &  2.9$\times10^{-126}$ & 6.4$\times10^{-123}$ &
 1.1$\times10^{-120}$ \\
\hline
\end{tabular}
\caption[]{ Minimum values of the electron mass for
convergence of $\alpha^{eff}(M_Z)$, for different values of $M_Z$, in the
Standard Electroweak Model. $M_Z = \sqrt{\alp}\upsilon'$.}    
\end{center}
\end{table}
\newpage
\vspace*{4cm}
\begin{figure}[htbp]
\begin{center}
\mbox{\epsfig{file=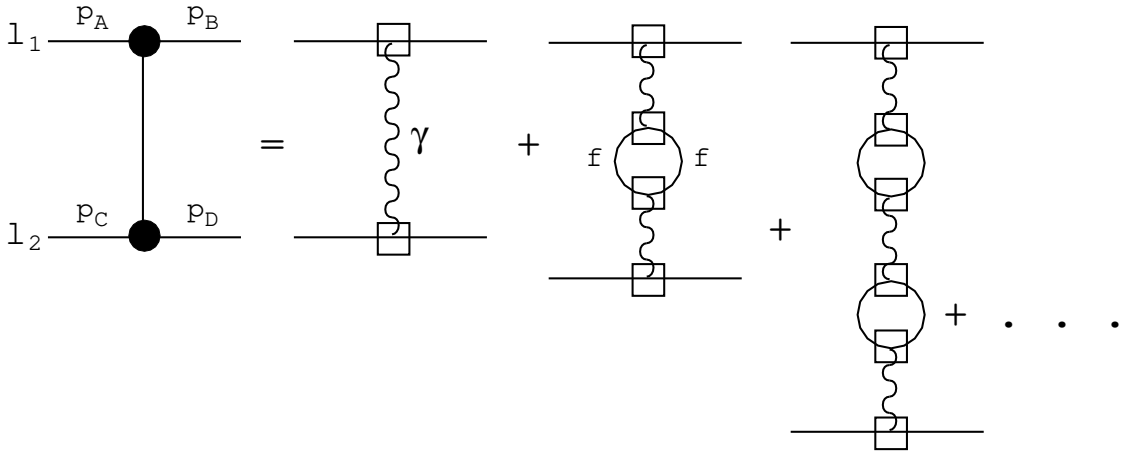,height=6cm}}
\caption{ Dyson sum of the 1-loop FVPL generating the effective QED
coupling constant in the process $l_1l_2 \rightarrow l_1l_2$. Open
square boxes: renormalised coupling constant with arbitary renormalisation
scale. Solid circles: renormalised coupling constant at the physical 
scale Q. Solid vertical line: dressed photon propagator.   }
\label{fig-fig1}
\end{center}
 \end{figure}  
\begin{figure}[htbp]
\begin{center}\hspace*{-0.5cm}\mbox{
\epsfysize10.0cm\epsffile{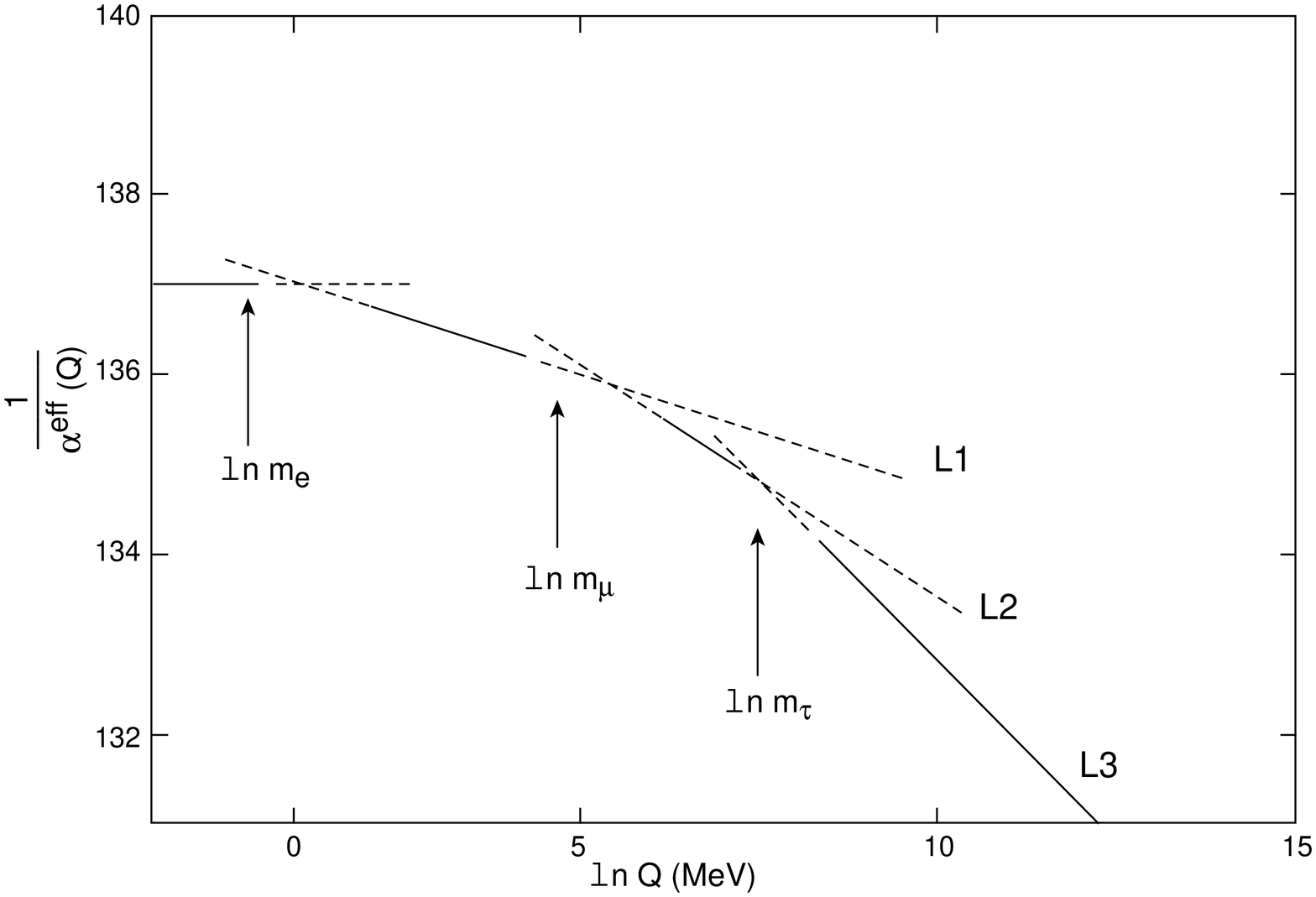}}
\caption{ $[\alpha^{eff}(Q)]^{-1}$ versus lnQ (MeV), showing the
decoupling of the FVPL due to $\tau$, $\mu$, $e$. Quark contributions are 
neglected. The solid lines indicate regions where the asymptotic formula
 (2.8) is valid.   }
\label{fig-fig2}
\end{center}
 \end{figure}  
\begin{figure}[htbp]
\begin{center}\hspace*{-0.5cm}\mbox{
\epsfysize10.0cm\epsffile{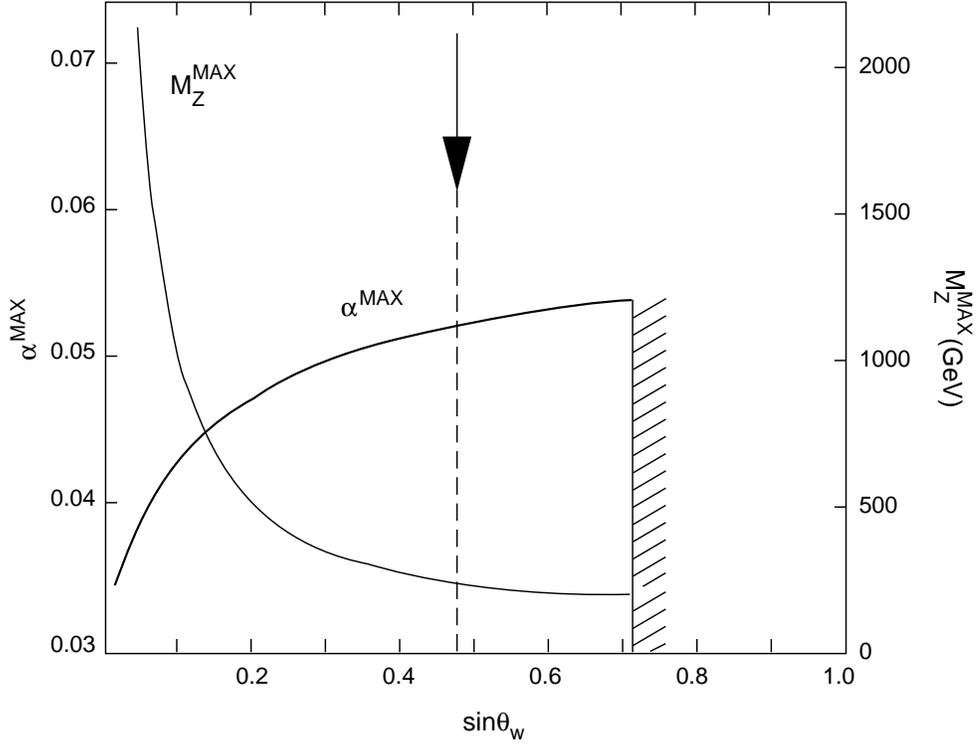}}
\caption{ Upper limits on the fine structure constant, $\alpha$, and the Z mass, 
 $M_Z$ in the Standard Electroweak Model as a function of $\sin \theta_W$
 (see text). $\upsilon = 252.2$ GeV is assumed. The vertical arrow shows
 the experimental value of $\sin \theta_W$. The cross-hatched theoretical bound 
corresponds to the minimum value of $M_Z^{MAX}$ 
and the maximum value of $\alpha^{MAX}$,  obtained when
 $\sin \theta_W = 1/\sqrt{2}$.}

\label{fig-fig3}
\end{center}
 \end{figure}  
\begin{figure}[htbp]
\begin{center}\hspace*{-0.5cm}\mbox{
\epsfysize10.0cm\epsffile{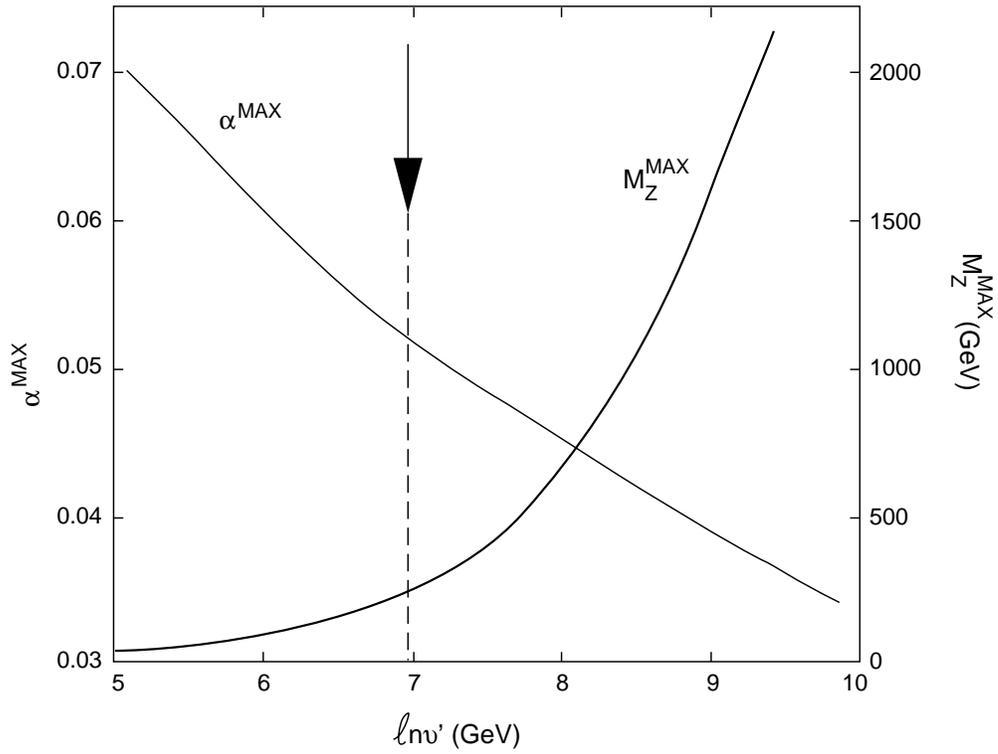}}
\caption{ Upper limits on the fine structure constant $\alpha$ and the Z mass $M_Z$ 
 in the Standard Electroweak Model as a function of $\ln \upsilon'$ (GeV)
 (see text). The vertical arrow shows the experimental value of
 $\ln \upsilon'$.} 
\label{fig-fig4}
\end{center}
 \end{figure}  
\end{document}